# On the relationship of the 27-day variations of the solar wind velocity and galactic cosmic ray intensity in minimum epoch of solar activity

M.V. Alania (1,2), R. Modzelewska (1), A. Wawrzynczak (3)

1. Institute of Mathematics and Physics, University of Natural Sciences and Humanities in Siedlce, 3 Maja 54, 08-110 Siedlce, Poland

2. Institute of Geophysics, Georgian Academy of Sciences, Tbilisi, Georgia

3. Institute of Computer Science, University of Natural Sciences and Humanities in Siedlce, 3 Maja 54, 08-110 Siedlce, Poland

*Fax: +48* 025-644-20-45

alania@uph.edu.pl; renatam@uph.edu.pl; awawrzynczak@uph.edu.pl*;*

## Abstract

We study the relationship of the 27-day variation of the galactic cosmic ray intensity with similar changes of the solar wind velocity and the interplanetary magnetic field based on the experimental data for the Bartels rotation period # 2379 of 23 November 2007 – 19 December 2007. We develop a three dimensional (3-D) model of the 27-day variation of galactic cosmic ray intensity based on the heliolongitudinally dependent solar wind velocity. A consistent, divergence-free interplanetary magnetic field is derived by solving Maxwell's equations with a heliolongitudinally dependent 27-day variation of the solar wind velocity reproducing in situ observations. We consider two types of 3-D models of the 27-day variation of galactic cosmic ray intensity - (1) with a plane heliospheric neutral sheet, and (2)- with the sector structure of the interplanetary magnetic field. The theoretical calculation shows that the sector structure does not influence significantly on the 27-day variation of galactic cosmic ray intensity as it was shown before based on the experimental data. Also a good agreement is found between the time profiles of the theoretically expected and experimentally obtained first harmonic waves of the 27-day variation of the galactic cosmic ray intensity (correlation coefficient equals 0.98 ± 0.02). The expected 27-day variation of the galactic cosmic ray intensity is inversely correlated with the modulation parameter $\zeta$ (correlation coefficient equals -0.91 ± 0.05) which is proportional to the product of the solar wind velocity V and the strength of the interplanetary magnetic field B ($\zeta \sim VB$). The high anticorrelation between these quantities indicates that the predictable 27-day variation of the galactic cosmic ray intensity mainly is caused by this basic modulation effect.

## 1. Introduction

Richardson, Cane and Wibberenz (1999) showed, that the recurrent 27-day variation of solar wind parameters, as well as of the galactic cosmic ray (GCR) intensity, are larger for positive (A>0) polarity epochs of solar magnetic cycles, than for negative ones (A<0). Previously, Alania et al., (2001a, 2001b); Gil and Alania (2001); Vernova et al., (2003); Iskra et al., (2004) demonstrated that the amplitudes of the 27-day variation of the GCR intensity obtained from neutron monitors are greater in the minimum epochs of solar activity for A>0 than for A<0. Recently, we demonstrated (Alania et al., 2005; Alania, Gil and Modzelewska 2008a, 2008b; Gil et al., 2005) that also the amplitudes of the 27-day variation of the GCR anisotropy at solar minimum are greater when A>0 than when A<0.

Many of the papers (Gil and Alania, 2001; Burger and Hitge, 2004; Alania et al., 2005; Gil et al., 2005; Alania, Gil and Modzelewska 2008a, 2008b; Burger et al., 2008) aimed to explain a dependence of the 27-day variation of the cosmic ray intensity on the A>0 and A<0 polarity epochs in terms of drift effect for the same heliolongitudinal changes of the solar wind velocity in different polarity epochs. At the same time, the role of different ranges of the heliolongitudinal changes of the solar wind velocity in the A>0 and A<0 epochs was not considered. The 3-D models with heliolongitudinal dependent solar wind velocity with consequent changes of the frozen interplanetary magnetic field (IMF) were presented by Kota and Jokipii (1991; 2001) followed after the first 3-D model (Kota and Jokipii, 1983). Model of Kota and Jokipii (1991) was the first global simulations of the modulation of GCR by 3-D solar wind with corotating interaction regions (CIR), though there was not considered an expected difference of the amplitudes of the 27-day variations of the GCR intensity in the A>0 and A<0 polarity epochs. Later, Kota and Jokipii (2001) developed new time dependent 3-D model including multiple heliospheric neutral sheet (HNS) characterizing high solar activity to cover more complex structures of the HNS and CIRs. It was demonstrated that the recurrent variation (27-days) of the GCR intensity is larger in the A>0 polarity epoch than in the A<0 epoch in qualitative agreement with the results of Richardson, Cane and Wibberenz (1999).

3-D models developed by Kota and Jokipii (1983; 1991; 2001) show that in the global modulation model of GCR, the large-scale structure is controlled by drift effects in conjunction with diffusion, convection and energy change, but the small-scale structure is caused by diffusive effects in the transient and corotating structures. Alania, Gil and Modzelewska (2008a), Gil, Alania and Modzelewska (2008) demonstrated that for the polarity dependent amplitudes of the 27-day variations of the GCR intensity are responsible, among the equally important drift effects, the heliolongitudinally dependent solar wind velocity with various ranges in the A>0 and A<0 polarity epochs. We showed (Alania, Modzelewska and Wawrzynczak, 2010) that to explain features of the 27-day variation of the GCR intensity by three dimensional (3-D) modelling, there should be taken into account a consistent, divergence-free IMF derived from Maxwell's equations with the heliolongitudinally dependent solar wind velocity reproducing in situ observations. Our general statement is that a proposed model should be able to describe not only average properties of the 27-day variation of the GCR intensity, but it must be adjustable to explain the behavior of the GCR intensity during any individual Bartels rotation (BR) period (27-days) taking into account in situ measurements of the solar wind velocity and the components of the IMF.

Our aim in this paper is threefold: (1) to study relationships of the 27-day variation of the GCR intensity, solar wind velocity and the IMF based on the experimental data for Sun's BR # 2379 of 23 November 2007–19 December 2007, (2) to compose a 3-D model of the 27-day variation of the GCR intensity including the solar wind velocity depending on heliolongitudes (reproducing in situ measurements) and corresponding divergence-free IMF derived from Maxwell's equations, and (3) to carry out a comparison and analysis of the results obtained based on the experimental data and modeling.

## 2. Data analysis for Bartels Rotation # 2377-2381 of 30 September 2007– 11 February 2008

Figure 1 presents temporal changes of the daily solar wind velocity [OMNI], GCR intensity from the Kiel neutron monitor, and radial $B_x$, azimuthal $B_y$ and latitudinal $B_z$ components of the IMF [OMNI] for five successive BR periods #2377-2381 of 30 September 2007– 11 February 2008. Figure 1 shows that the quasi periodic changes ~ 27 days are more or less recognizable in all parameters except the $B_z$ component. Due to extremely small values of the $B_z$ component its contribution in the magnitude of the IMF is negligible, so, in this paper, we neglect its role in further study. Our aim is to compose a 3-D model of the 27-day variation capable to explain the behavior of the GCR intensity during single BR period. Generally, for analysis one can choose arbitrarily any BR period from five successive BR periods # 2377-2381. However, we suppose that to find single-valued reliable relationship of the recurrent changes of the GCR intensity variation with the similar changes of the solar activity and solar wind parameters, it is reasonable to analyze a BR period with clearer pronunciations of these changes.

Table 1 presents the correlation coefficients, *r,* between changes of the GCR intensity variations and solar wind velocity for each individual BR period # 2377-2381 and for the whole period of 30 September 2007– 11 February 2008. Table 1 shows a clear anticorrelation for each BR periods # 2377-2381 with the exception of last BR period #2381. In general, this BR period # 2381 is the reason that the correlation coefficient for whole five BR periods # 2377-2381 is relatively small, r=-0.53±0.004, while without the BR period # 2381 the correlation coefficient is
r=-0.59±0.004. To compose a 3-D model of the 27-day variation capable to explain the behavior of the GCR intensity during a single BR period, we have chosen the BR period #2379 of 23 November 2007 – 19 December 2007 (black box in Figure 1) with the highest correlation coefficient, r=-0.81±0.01.

Figure 2ab shows daily data (points) and the first harmonic waves of the 27-day variations (dashed lines) of the solar wind velocity (Figure 2a), and the GCR intensity (Figure 2b) during the BR # 2379 period (23 November 2007 – 19 December 2007). The first harmonic wave of the 27-day variation (dashed line) of the solar wind velocity (Figure 2a) is expressed as:

$$V_r = V_0(1 + \alpha \sin(\varphi - \varphi_0)), \tag{1}$$

where $V_0 = 442$ km s$^{-1}$, $\alpha = -0.3$, $\varphi_0 = 1.4$ (in radians).

The correlation coefficient between the changes of the experimental daily data (points) of the solar wind velocity (Figure 2a) and the GCR intensity (Figure 2b) is negative and equals -0.81 ± 0.01, while the correlation coefficient between the solar wind velocity and the GCR intensity represented by the first harmonic waves of the 27-day variations (dashed lines, Figure 2ab) is higher and equals - 0.99 ± 0.02.

Thus, the experimental data show that a role of the first harmonic of the 27-day variation of the solar wind velocity is decisive in the creation of the first harmonic of the 27-day variation of the GCR intensity during BR # 2379 (23 November 2007 – 19 December 2007) in the minimum epoch of solar activity. Similar dependence between first harmonic of the 27-day variation of the solar wind velocity and the first harmonic of the 27-day variation of the GCR intensity was previously found for other periods (Modzelewska et al., 2006; Alania, Gil and Modzelewska, 2008a).

## 3. Model of the 27-Day variation of the GCR intensity

We model the 27-day variation of the GCR intensity based on the Parker's transport equation (Parker, 1965):

$$\frac{\partial f}{\partial t} = \nabla \cdot \left(K_{ij}^S \cdot \nabla f\right) - (v_d + V) \cdot \nabla f + \frac{1}{3}(\nabla \cdot V)\frac{\partial f}{\partial \ln R} \tag{2}$$

Where $f$ and $R$ are omnidirectional distribution function and rigidity of cosmic ray particles, respectively; $V$ – solar wind velocity, $t$ - time, $v_d$ is the drift velocity $v_{d,i} = \partial K_{ij}^A / \partial x_j$ (Jokipii, Levy and Hubbard, 1977), $K_{ij}^S$ and $K_{ij}^A$ are the symmetric and antisymmetric parts of anisotropic diffusion tensor $K_{ij} = K_{ij}^S + K_{ij}^A$ of GCR. The omnidirectional distribution function $f(\vec{r}, \vec{p})$ is related to the differential intensity I, as $I = p^2 f$ (p is momentum of cosmic ray particle).

We implement in the model of the 27-day variation of the GCR intensity (Equation (2)) the first harmonic of the 27-day variation of the solar wind velocity representing in situ measurements of the BR # 2379 period (Equation (1)) and corresponding components $B_r$ and $B_\varphi$ of the IMF obtained as the solutions of Maxwell's equations for the variable solar wind velocity (Equation (1)).

### 3.1. Model of the IMF for the variable solar wind velocity and numerical solution of Maxwell's equation

Maxwell's equation for infinite conductivity plasma can be written (e.g. Maxwell, 1865; Parker, 1963; Jackson, 1998; Hanslmeier, 2002):

$$\frac{\partial \vec{B}}{\partial t} = \nabla \times (\vec{V} \times \vec{B}) \tag{3}$$

where $\vec{V}$ is the solar wind velocity, $\vec{B}$ –magnetic induction. To find the components of the IMF we have to solve the system consisting of the vector Equation (3) and the scalar equation describing divergence free IMF:

$$\nabla \cdot \vec{B} = 0 \qquad (4)$$

The system of scalar equations for the components $(B_r, B_\theta, B_\varphi)$ of the IMF and components $(V_r, V_\theta, V_\varphi)$ of the solar wind velocity corresponding to system of Equation (3) together with the requirement given by Equation (4) can be rewritten in heliocentric spherical $(r, \theta, \varphi)$ coordinate system corotating with the Sun, as:

$$\begin{cases} \dfrac{\partial B_r}{\partial t} = \dfrac{1}{r^2 \sin\theta}\left[\dfrac{\partial}{\partial \theta}[(V_r B_\theta - V_\theta B_r)r\sin\theta] - \dfrac{\partial}{\partial \varphi}[(V_\varphi B_r - V_r B_\varphi)r]\right] & (5a) \\[4pt]
\dfrac{\partial B_\theta}{\partial t} = \dfrac{1}{r\sin\theta}\left[\dfrac{\partial}{\partial \varphi}(V_\theta B_\varphi - V_\varphi B_\theta) - \dfrac{\partial}{\partial r}[(V_r B_\theta - V_\theta B_r)r\sin\theta]\right] & (5b) \\[4pt]
\dfrac{\partial B_\varphi}{\partial t} = \dfrac{1}{r}\left[\dfrac{\partial}{\partial r}[(V_\varphi B_r - V_r B_\varphi)r] - \dfrac{\partial}{\partial \theta}(V_\theta B_\varphi - V_\varphi B_\theta)\right] & (5c) \\[4pt]
\dfrac{1}{r}\dfrac{\partial}{\partial r}(r^2 B_r) + \dfrac{1}{\sin\theta}\dfrac{\partial}{\partial \theta}(\sin\theta B_\theta) + \dfrac{1}{\sin\theta}\dfrac{\partial}{\partial \varphi}B_\varphi = 0 & (5d) \end{cases}$$

Our aim in this paper is to describe 27-day variation of the GCR intensity during one BR period. For this purpose we suggest that the 3-D model of the 27-day variation of the GCR intensity reflects processes corresponding to the instant state of the heliosphere during one BR period (27-days). So, the heliolongitudinal changes of the solar wind velocity, $B_x$, $B_y$ and $B_z$ components of the IMF can be considered as a quasi stationary for one BR period of the Sun, i.e. the spatial distribution of the GCR density is determined by the longitudinally dependent, but steady-state parameters - the solar wind velocity, $B_x$, $B_y$ and $B_z$ components of the IMF. For that reason, we accept that $\dfrac{\partial B_r}{\partial t} = 0$, $\dfrac{\partial B_\theta}{\partial t} = 0$, $\dfrac{\partial B_\varphi}{\partial t} = 0$ in Equations (5a)-(5c). Also, we accept that average value of the heliolatitudinal component of the solar wind velocity $V_\theta$ equals zero. Then the system of Equations (5a)-(5d) can be reduced, as

$$\begin{cases} \sin\theta V_r \dfrac{\partial B_\theta}{\partial \theta} + \sin\theta B_\theta \dfrac{\partial V_r}{\partial \theta} + \cos\theta V_r B_\theta - V_\varphi \dfrac{\partial B_r}{\partial \varphi} - B_r \dfrac{\partial V_\varphi}{\partial \varphi} + V_r \dfrac{\partial B_\varphi}{\partial \varphi} + B_\varphi \dfrac{\partial V_r}{\partial \varphi} = 0 & (6a) \\[4pt]
V_\varphi \dfrac{\partial B_\theta}{\partial \varphi} + B_\theta \dfrac{\partial V_\varphi}{\partial \varphi} + r\sin\theta V_r \dfrac{\partial B_\theta}{\partial r} + r\sin\theta B_\theta \dfrac{\partial V_r}{\partial r} + \sin\theta V_r B_\theta = 0 & (6b) \\[4pt]
rB_r \dfrac{\partial V_\varphi}{\partial r} + rV_\varphi \dfrac{\partial B_r}{\partial r} + V_\varphi B_r - V_r B_\varphi - rV_r \dfrac{\partial B_\varphi}{\partial r} - rB_\varphi \dfrac{\partial V_r}{\partial r} + B_\theta \dfrac{\partial V_\varphi}{\partial \theta} + V_\varphi \dfrac{\partial B_\theta}{\partial \theta} = 0 & (6c) \\[4pt]
\dfrac{\partial B_r}{\partial r} + \dfrac{2}{r}B_r + \dfrac{ctg\theta}{r}B_\theta + \dfrac{1}{r}\dfrac{\partial B_\theta}{\partial \theta} + \dfrac{1}{r\sin\theta}\dfrac{\partial B_\varphi}{\partial \varphi} = 0. & (6d) \end{cases}$$

The latitudinal component $B_\theta$ of the IMF is very feeble for the period to be analyzed, so we can assume that $B_\theta = 0$, so further we consider Parker's spiral 2-D model of the IMF. This assumption straightforwardly leads (from Equation (6a)) to the relationship between $B_r$ and $B_\varphi$, as $B_\varphi = B_r \dfrac{V_\varphi}{V_r}$. Then Equation (6d) with respect to the radial component $B_r$ has a form:

$$A_1 \dfrac{\partial B_r}{\partial r} + A_2 \dfrac{\partial B_r}{\partial \varphi} + A_3 B_r = 0. \qquad (7)$$

The coefficients $A_1$, $A_2$ and $A_3$ depend on the radial $V_r$ and heliolongitudinal $V_\varphi$ components of the solar wind velocity $\vec{V}$.

We take the corotational velocity, as well:

$$V_\varphi = -\Omega r \sin\theta, \qquad (8)$$

where $\Omega$ is the angular velocity of the Sun.

Taking into account the Equations (1) and (8) the coefficients $A_1$, $A_2$ and $A_3$ in Equation (7) are:

$$A_1 = 1 \quad A_2 = -\frac{\Omega}{V_r} \quad A_3 = \frac{2}{r} + \frac{\Omega}{V_r^2}\frac{\partial V_r}{\partial \varphi}.$$

The kinematical model of the IMF with variable solar wind velocity has some limitations, especially it can be applied until some radial distance, while at large radii the faster wind would overtake the previously emitted slower one. To exclude an intersection of the IMF lines the heliolongitudinal asymmetry of the solar wind velocity takes place only up to the distance of ~ 8 AU and then $V = 400$ km s$^{-1}$ throughout the heliosphere; for that reason, behind 8 AU the standard Parker's magnetic field is applied. At this distance the IMF is weak and almost tangential, so is not sensitive at all to transit from the heliolongitudinal asymmetry region of the heliosphere to the Parker's type heliosphere region. We solve the first order linear partial differential Equation (7) by numerical method. For this purpose, Equation (7) was reduced to the algebraic system of equations using a difference scheme method (e.g., Kincaid and Cheney, 2002), as

$$A_1 \frac{B_r[i+1,j,k]-B_r[i,j,k]}{\Delta r} + A_2 \frac{B_r[i,j,k+1]-B_r[i,j,k]}{\Delta \varphi} + A_3 B_r[i,j,k] = 0, \qquad (9)$$

where, i=1,2,…, I; j=1,2,…, J; k=1,2,…, K are steps in radial distance, vs. heliolatitude and heliolongitude, respectively. Then Equation (9) was solved by the iteration method with the boundary condition near the Sun $B_r[i,j,k]\big|_{r=r_1} = const$. In the considered case $r_1$ =0.1 AU which corresponds to the radial distance, where an acceleration of solar wind takes place in the minimum epochs of solar activity, $B_r[i,j,k]\big|_{r=r_1} = A \cdot 1400\ nT$, where A is a signed constant; positive when the field in the northern hemisphere is directed away from the Sun, and negative when it is directed toward the Sun.

The choice of these boundary conditions was stipulated by requiring agreement of the solutions of Equation (9) with the in situ measurements of the $B_r$ and $B_\varphi$ components of the IMF at the Earth orbit.

Figures 3-5 show results of the solution of Equation (9) for the $B_r$ and $B_\varphi$ components of the IMF. For comparison Figure 6ab presents in situ measurements of the heliolongitudinal changes of the $B_x$ and $B_y$ components of the IMF (points) and the first harmonic waves of the 27-day variations (dashed lines) in the period of the BR #2379. Figures 5ab and 6ab show that the character of the 27-day waves of the observed $B_x$ and $B_y$ components of the IMF coincide with the solutions of Maxwell's equations. However, as it was expected, Maxwell's equation due to its specific solutions (connected with the boundary conditions) does not provide a sector structure, observed in the experimental data due to non coincidence of the magnetic dipole and heliographic axes of the Sun (Schulz, 1973). The sign of B is adjusted later in accordance with the sector structure observed at 1 AU.

**3.2 Model of the 27-day variation of the GCR intensity and its numerical solution.**

The proposed 3-D model of the 27-day variation of the GCR intensity corresponds to the instant state of the heliosphere during one BR period (27-days) and it can be considered as a stationary one; then an assumption $\frac{\partial f}{\partial t} = 0$ is acceptable in Equation (2).

For the dimensionless variables $f^* = \frac{f}{f_0}$, and $r = \frac{r^*}{r_0}$, where $f_0$ is density in the local interstellar medium, $r_0$- size of modulation region, in the spherical coordinate system $(r, \theta, \varphi)$ Equation (2) can be written as:

$$\Lambda_1 \frac{\partial^2 f^*}{\partial r^2} + \Lambda_2 \frac{\partial^2 f^*}{\partial \theta^2} + \Lambda_3 \frac{\partial^2 f^*}{\partial \varphi^2} + \Lambda_4 \frac{\partial^2 f^*}{\partial r \partial \theta} + \Lambda_5 \frac{\partial^2 f^*}{\partial r \partial \varphi}$$
$$+ \Lambda_6 \frac{\partial^2 f^*}{\partial \theta \partial \varphi} + \Lambda_7 \frac{\partial f^*}{\partial r} + \Lambda_8 \frac{\partial f^*}{\partial \theta} + \Lambda_9 \frac{\partial f^*}{\partial \varphi} + \Lambda_{10} f^* + \Lambda_{11} \frac{\partial f^*}{\partial R} = 0. \quad (10)$$

Where coefficients $\Lambda_1, \Lambda_2, ..., \Lambda_{11}$ depend on spatial coordinates $(r, \theta, \varphi)$ and rigidity R.

The Equation (10) was transformed to the algebraic system of equations using the implicit finite difference scheme method and then solved by the Gauss-Seidel method of iteration as in papers published elsewhere (e.g., Alania, 2002; Iskra et al., 2004; Modzelewska et al., 2006; Wawrzynczak and Alania, 2008).

The parallel diffusion coefficient $\kappa_\parallel$ changes versus the spatial spherical coordinates $(r, \theta, \varphi)$ and rigidity R of GCR particles as,

$$\kappa_\parallel = \kappa_0 \kappa_1(r) \kappa_2(R), \quad (11)$$

where $\kappa_0 = \frac{\lambda_0 v}{3} = 4 \times 10^{22}$ cm$^2$s$^{-1}$, $v$ is the velocity of GCR particles, and $\lambda_0$ - the mean free path of GCR particles ($\lambda_0 = 4 \times 10^{12}$ cm for the rigidity of 10GV); $\kappa_2(R) = (R/1GV)^{0.8}$; $\kappa_1(r) = 1 + \alpha_0 r$, where $\alpha_0 = 0.5$, and r is a radial distance in AU; the term $\alpha_0 r$ stand for a dependence of the parallel diffusion coefficient $\kappa_\parallel$ on the magnitude B of the IMF; the value $\alpha_0 = 0.5$ is chosen by such way that a dependence $B \propto \frac{1}{r}$ provides a consistence with normalized parallel diffusion coefficient $\kappa_\parallel \approx 4 \times 10^{23}$ cm$^2$s$^{-1}$ for GCR particle of 10GV rigidity at the Earth orbit. The ratios of $\beta$ and $\beta_1$ of the perpendicular $\kappa_\perp$ and drift $\kappa_d$ diffusion coefficients to the parallel diffusion coefficient $\kappa_\parallel$ of the GCR particles are given in standard form, $\beta = \frac{\kappa_\perp}{\kappa_\parallel} = (1 + \omega^2 \tau^2)^{-1}$ and $\beta_1 = \frac{\kappa_d}{\kappa_\parallel} = \omega \tau (1 + \omega^2 \tau^2)^{-1}$, where $\omega \tau = 300 \, B \lambda_0 R^{-1}$, B is the strength of the IMF. These expressions of $\beta$ and $\beta_1$ is acceptable for the scattering of the GCR particles to which neutron monitors and ground muon telescopes respond and it is in accord with the quasi linear theory (Jokipii, 1971; Shalchi, 2009).

The vector $\vec{B}$ of the IMF may be written (Jokipii and Kopriva, 1979):

$$\vec{B} = \left(1 - 2H(\theta - \theta')\right)\left(B_r \vec{e}_r + B_\varphi \vec{e}_\varphi\right), \quad (12)$$

where H is the Heaviside step function changing the sign of the global magnetic field in each hemisphere, $\vec{e}_r$ and $\vec{e}_\varphi$ are the unite vectors directed along the component $B_r$ and $B_\varphi$ of the IMF and $\theta'$ corresponds to the heliolatitudinal position of the HNS.

Subsequent papers (Alania et al., 2001b; Alania et al., 2005; Gil et al., 2005) show the results that the amplitudes of the 27-day variations of the GCR intensity and anisotropy do not depend on the tilt angles of the HNS during the 11-year cycle of solar activity. It is acceptable to use the planar HNS in modeling of the 27-day variation of the GCR intensity. Nevertheless, as the sector structure of the IMF is observed, it is of interest to take into account waviness of the HNS in modeling of the 27-day variation of the GCR intensity. We implement the sector structure in the model of the 27-day variation of the GCR intensity using the formula suggested by Jokipii and Thomas (1981); the formula adjusting to the period of BR # 2379 can be expressed, as:

$$\theta' = \frac{\pi}{2} + \delta \sin\left(\varphi + \frac{r\Omega}{V_r} + 0.87\right), \quad (13)$$

where $\delta = 15^0$ is the observed tilt angle during BR period # 2379 and the value 0.87 (in radian) - shifting angle of the HNS with respect to the heliolongitude corresponding to the BR period # 2379.

The neutral sheet drift was taken into account according to the boundary condition method (Jokipii and Kopriva, 1979), when the delta function at the HNS is a consequence of the abrupt change in sign of the IMF.

The model of the 27-day variation of the GCR intensity (Equation (10)) incorporates the first harmonic of the 27-day variation of the solar wind velocity (Equation (1)) corresponding to in situ measurements of the BR # 2379 period and the components $B_r$ and $B_\varphi$, and the magnitude $B = \sqrt{B_r^2 + B_\varphi^2}$ of the IMF obtained from the numerical solution of Equation (9).

The components $B_r$ and $B_\varphi$ appear in the spiral angle, $\psi = \arctan\left(-\frac{B_\varphi}{B_r}\right)$, i.e. in the principal directions of the anisotropic diffusion coefficient; while the magnitude $B = \sqrt{B_r^2 + B_\varphi^2}$ is explicitly introduced via the value of the parallel diffusion coefficient, $\kappa_\parallel$ (Equation (11)).

The model of the 27-day variation of the GCR intensity (Equation (10)) was solved for two cases: (1) for the planar HNS and (2) for the wavy HNS.

As it is seen from Figure 6ab, experimental data of the $B_x$ and $B_y$ components of the IMF show sector structure, while as was mentioned above these components found as the solutions of Maxwell's equation (Figure 5ab) do not provide sector structure. However, we implement in the numerical model of the 27-day variation of the GCR intensity the hybrid scheme which includes $B_r$ and $B_\varphi$ components obtained as the solutions of Maxwell's equation and sector structure according to formulas (12) and (13) corresponding to in situ measurements for BR # 2379 (Figure 7ab). Figure 7ab shows the heliolongitudinal changes of the $B_r$ and $B_\varphi$ components of the IMF obtained as the solution of Maxwell's equations, taking into account sector structure of the IMF (points) and its first harmonic waves of the 27-day variations (dashed lines). Comparing Figures 6ab and 7ab one can find that the 27-day waves of the observed $B_x$ and $B_y$ components of the IMF coincide with the theoretical $B_r$ and $B_\varphi$ components of the IMF included in the model of the 27-day variation of the GCR intensity.

Figure 8 shows the experimental data of Kiel neutron monitor (points) for the one BR # 2379 period of 23 November 2007–19 December 2007 and its first harmonic wave (dotted line); also in this figure are presented heliolongitudinal changes of the expected first harmonic waves of the 27-day variation of the GCR intensity for rigidity 10GV at the Earth orbit obtained as a solution of two models: (1) with the planar HNS (dashed line) and (2) with the sector structure of the IMF (solid line). Comparing the results of two different models of the 27-day variation of the GCR intensity one can conclude that a consideration of the sector structure of the IMF does not cause significant effect. Figure 8 shows an acceptable agreement of the expected changes of the amplitudes of the 27-day variation of the GCR intensity obtained by theoretical modeling with the experimental data.

It should be pointed out that in the proposed 3-D model of the 27-day variation of the GCR intensity an important and dominant role is played by the 27-day changes of the solar wind velocity together with the diffusion coefficient K of the GCR particles, which depends, among other parameters, on the magnitude B of the regular IMF, as K∝1/B. Certainly, the transport equation in this case explicitly contains the modulation parameter ζ which is proportional to the product of the solar wind velocity V and the strength B of the regular IMF (ζ ~ VB). This is a hidden effect of the electric field in transport equation being responsible, in general, for the convection-diffusion propagation of GCR (Parker, 1965; Gleeson and Axford, 1967), and for recurrent changes of GCR, as well (Jokipii and Kota, 1991). Figure 9 demonstrate the changes of the parameter ζ ~ VB (dashed line) at the Earth orbit for the solar wind velocity V given by the Equation (1) and magnitude B of the IMF obtained as the numerical solution of Equation (9), and the expected 27-day wave of the GCR intensity for rigidity 10 GV obtained as a solution of the 3-D model (solid line) with wavy HNS (Equation (10)). Figure 9 shows that there is a remarkable negative correlation between the modulation parameter ζ and expected 27-day wave of the GCR intensity, r = -0.91 ± 0.05. An important role of the modulation parameter ζ (determined by the solar wind velocity V and magnitude B of the IMF) is clearly manifested, for the energies much less than neutron monitor registered, in the observations at the distance of ~ 11 AU in interplanetary space by Voyager I (Burlaga et al., 1985). It was shown that when a turbulent interaction region moved past the spacecraft, the cosmic ray intensity (> 75 MeV/ nucleon particles) decreased by an amount proportional to the strength of the magnetic field B in the interaction region.

## 4. CONCLUSIONS

1. A three-dimensional model of the 27-day variation of GCR intensity is proposed based on the Parker's transport equation taking into account a consistent, divergence-free IMF derived from Maxwell's equations with the heliolongitudinally dependent solar wind velocity reproducing in situ observations.
2. We consider two types of 3-D models of the 27-day variation of GCR intensity: (1) with planar HNS, and (2) with the sector structure of the IMF. The theoretical calculation shows that the sector structure of the IMF does not influence significantly the 27-day variation of GCR intensity as it was also shown before based on the experimental data (Alania et al., 2001b; Alania et al., 2005; Gil et al., 2005). We show that the first harmonic waves of the theoretically expected 27-day variation of the GCR intensity (obtained as a solution of the 3 - D model) and Kiel neutron monitor data are in good agreement for BR period # 2379 of 23 November 2007 – 19 December 2007.
3. The expected 27-day variation of the GCR intensity is inversely correlated with the modulation parameter $\zeta$ being proportional to the product of the solar wind velocity V and the strength of the IMF B ($\zeta \sim VB$). This is a hidden effect of the electric field in transport equation being responsible, in general, for the convection-diffusion propagation of GCR, and for recurrent changes of GCR, as well.
4. High anticorrelation between theoretically expected 27-day variation of the GCR intensity and parameter $\zeta \sim VB$ shows that the 27-day variation of the GCR intensity takes place mainly due to this modulation effect in the minimum epoch of solar activity.

## Acknowledgments

The authors thank the referee whose valuable remarks and suggestions help us to improve the paper.

Authors thank the investigators of Kiel neutron monitor and SPDF OMNIWeb database.

FIGURE LEGEND

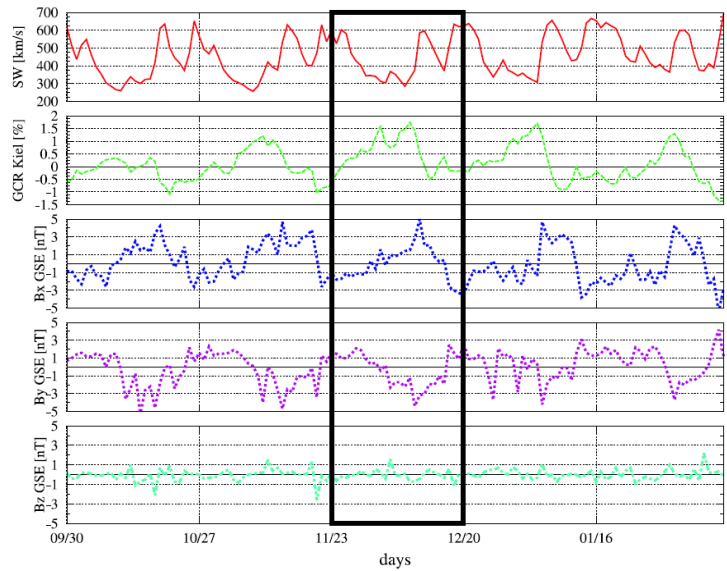

Figure. 1. Temporal changes of the daily solar wind velocity [OMNI, http://omniweb.gsfc.nasa.gov/index.html], GCR intensity from the Kiel neutron monitor [http://134.245.132.179/kiel/main.htm], and radial $B_x$, azimuthal $B_y$ and latitudinal $B_z$ components of the IMF [OMNI] for the period of 30 September 2007–11 February 2008 (BR # 2377-2381); black box designates BR #2379 of 23 November 2007- 19 December 2007.

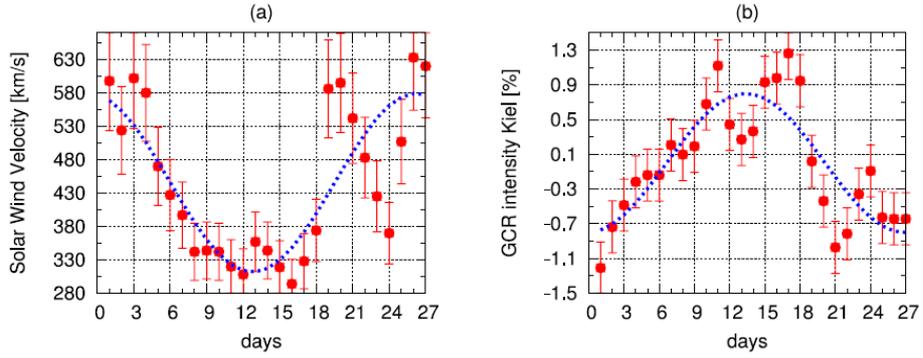

Figure 2ab. Temporal changes of daily data (points) and an approximation of the first harmonic waves of the 27-day variations (dashed lines) of the solar wind velocity (a), and the GCR intensity (b) during the period of 23 November 2007-19 December 2007 (BR # 2379).

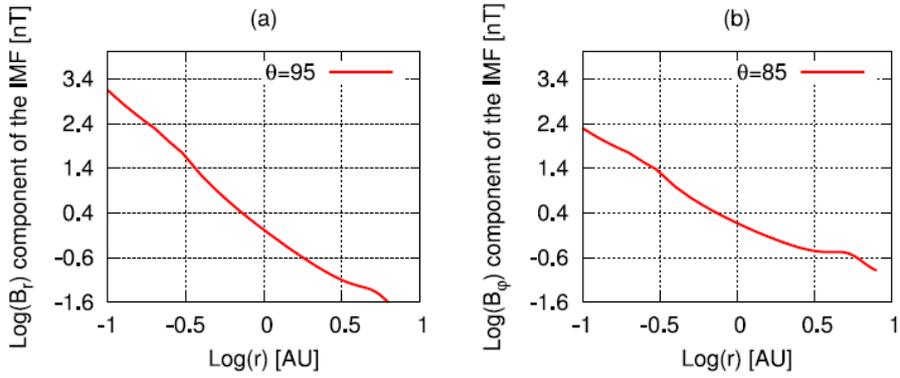

Figure 3ab Radial changes of the (a) $B_r$ and (b) $B_\varphi$ components of the IMF in log-log scale near the solar equatorial plane for fixed heliolongitude ($\varphi=180°$) for the solar wind velocity given by Equation (1).

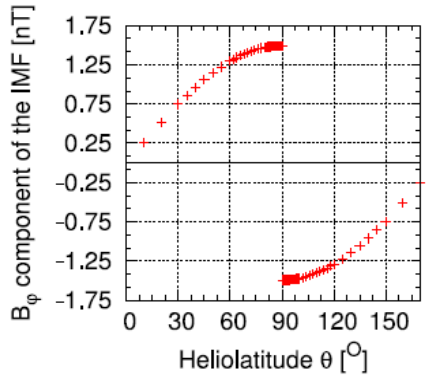

Figure 4. Heliolatitudinal changes of the $B_\varphi$ component of the IMF at 1 AU for the solar wind velocity given by Equation (1).

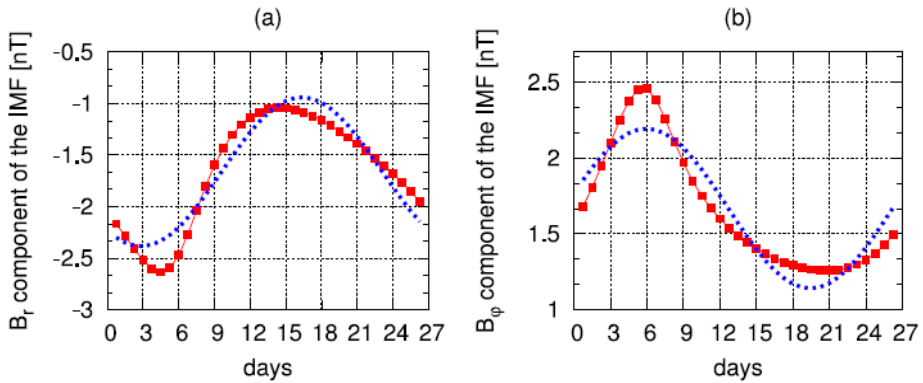

Figure 5ab. Heliolongitudinal changes of the (a) $B_r$ and (b) $B_\varphi$ components of the IMF (points) near Earth orbit for the solar wind velocity given by Equation (1) and the first harmonic waves of the 27-day variations (dashed lines).

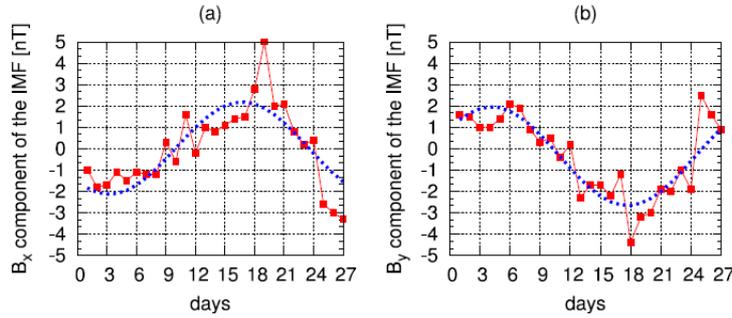

Figure 6ab. Heliolongitudinal changes of the (a) $B_x$ and (b) $B_y$ components of the IMF (points) measured at the Earth [OMNI] during the period of 23 November 2007-19 December 2007 (BR # 2379) and the first harmonic waves of the 27-day variations (dashed lines).

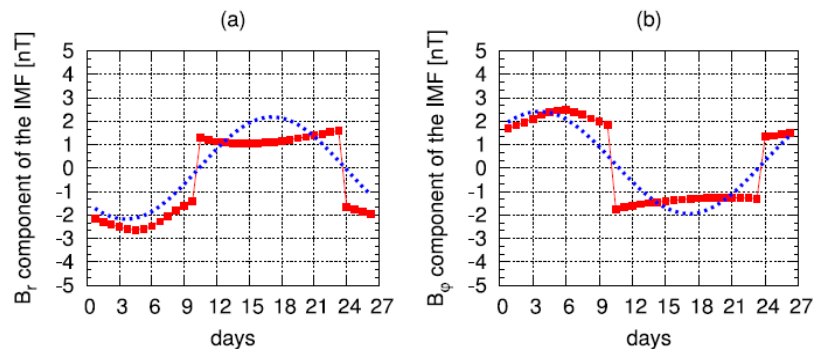

Figure 7ab. Heliolongitudinal changes of the (a) $B_r$ and (b) $B_\varphi$ components of the IMF (points) near Earth orbit obtained as the solution of Maxwell's equations for the solar wind velocity given by Equation (1) taking into account sector structure of the IMF and its first harmonic waves of the 27-day variations (dashed lines).

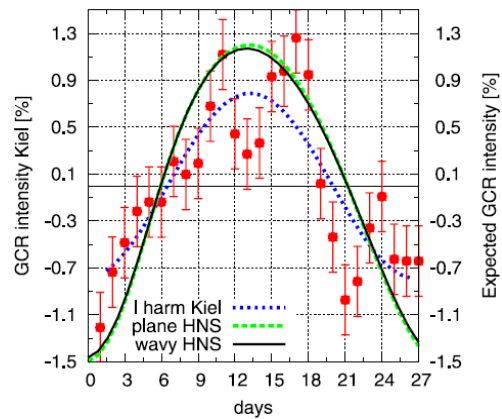

Figure 8. Heliolongitudinal changes of the expected first harmonic waves of the 27-day variation of the GCR intensity for rigidity 10GV at the Earth orbit obtained as a solution of two models: (1) - with the planar HNS (dashed line) and (2)- with the sector structure of the IMF (solid line) and temporal changes of GCR intensity by Kiel neutron monitor (points) and its first harmonic wave (dotted line) during the period of 23 November 2007-19 December 2007, BR # 2379.

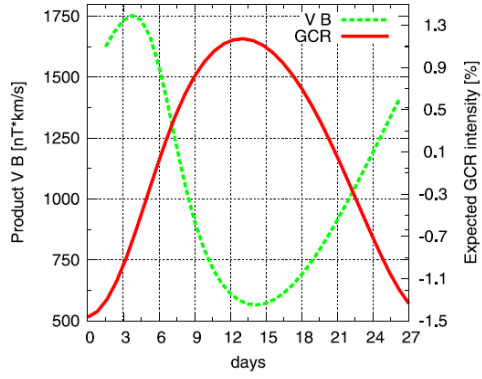

Figure 9. Heliolongitudinal changes of the parameter ζ~VB (dashed line) and the expected GCR intensity for rigidity 10GV at the Earth orbit obtained as a solution of the model with the sector structure of the IMF (solid line) during Sun's Bartels rotation # 2379 of 23 November 2007-19 December 2007.

Table 1 Correlation coefficients between the changes of the solar wind velocity and GCR intensity for the whole period (BR #2377 – 2381) and for each BR period separately.

| Bartels Rotation period | BRs #2377–2381 | BRs #2377–2380 | BR #2377 | BR #2378 | BR #2379 | BR #2380 | BR #2381 |
|---|---|---|---|---|---|---|---|
| Correlation coefficient | −0.53 ± 0.004 | −0.59 ± 0.004 | −0.76 ± 0.01 | −0.64 ± 0.02 | **−0.81 ± 0.01** | −0.64 ± 0.02 | −0.18 ± 0.02 |

Table 1. Correlation coefficients between the changes of the solar wind velocity and GCR intensity for the whole period (BR #2377-2381) and for each BR period separately.